\documentclass[reprint,superscriptaddress,amsmath,amssymb,aps,pra]{revtex4-1}

\usepackage{graphicx}
\usepackage{natbib}
\usepackage{hyperref}
\usepackage{cleveref}
\usepackage[caption=false,labelformat=empty]{subfig}
\usepackage{mathrsfs}
\usepackage{gensymb}

\hypersetup{colorlinks=true,citecolor=blue}

\newcommand{\Wcmsqd}{\mathrm{W}\text{cm}^{-2}}
\newcommand{\rmd}{\mathrm{d}}
\newcommand{\micron}{{\mu\mathrm{m}}}
\newcommand{\fs}{{\mathrm{fs}}}

\newcommand{\Ecrit}{E_\text{cr}}
\newcommand{\avg}[1]{{\left\langle {#1} \right\rangle}}
\newcommand{\abs}[1]{{\left| {#1} \right|}}
\newcommand{\waist}{w_0}

\begin{document}

\title{Radiation beaming in the quantum regime}

\author{T. G. Blackburn}
\email{tom.blackburn@physics.gu.se}
\affiliation{Department of Physics, University of Gothenburg, SE-41296 Gothenburg, Sweden}
\author{D. Seipt}
\affiliation{Center for Ultrafast Optical Science, University of Michigan, Ann Arbor, Michigan 48109, USA}
\author{S. S. Bulanov}
\affiliation{Lawrence Berkeley National Laboratory, Berkeley, California 94720, USA}
\author{M. Marklund}
\affiliation{Department of Physics, University of Gothenburg, SE-41296 Gothenburg, Sweden}

\date{\today}

\begin{abstract}
Classical theories of radiation reaction predict that the electron motion is confined
to the plane defined by the electron's instantaneous momentum and the force exerted by
the external electromagnetic field.
However, in the quantum radiation reaction regime, where the recoil exerted
by individual quanta becomes significant, the electron can scatter `out-of-plane',
as the photon is emitted into a cone with finite opening angle.
We show that Monte Carlo implementation of an angularly resolved emission rate
leads to substantially improved agreement with exact QED calculations
of nonlinear Compton scattering.
Furthermore, we show that the transverse recoil caused by this finite beaming,
while negligible in many high-intensity scenarios,
can be identified in the increase in divergence, in the plane perpendicular to the laser polarization and wavevector,
of a high-energy electron beam that interacts with a linearly polarized, ultraintense laser.
\end{abstract}

\maketitle

\section{Introduction}

Recent advances in the development of high-intensity lasers~\cite{sung.ol.2017,kiriyama.ol.2018,nakamura.ieee.2017}
and plasma-based accelerators~\cite{corde.nature.2015,adli.nature.2018,gonsalves.prl.2019} 
have made it possible to perform experiments on the interaction of charged particles
with ultraintense electromagnetic pulses in regimes previously
unexplored~\cite{cole.prx.2018,poder.prx.2018}.
Earlier experiments relied on conventional accelerator technology~\cite{bula.prl.1996,burke.prl.1997}. 
The processes studied belong to the field of high-intensity particle
physics~\cite{mourou.rmp.2006,marklund.rmp.2006,dipiazza.rmp.2012},
which combines quantum electrodynamics (QED) with the theory of strong
electromagnetic (EM) background fields~\cite{ritus.jslr.1985}.
Of particular significance is photon emission, because the recoil it
exerts can dominate the dynamics of electrons and positrons
in high-field environments, including neutron-star magnetospheres~\cite{harding.rpp.2006}
and laser-matter~\cite{ridgers.prl.2012,nakamura.prl.2012,stark.prl.2016}
or laser-laser~\cite{bell.prl.2008,fedotov.prl.2010,bulanov.prl.2010b,nerush.prl.2011}
interactions in next-generation
facilities~\cite{papadopoulos.hpl.2016,weber.mre.2017,gales.rpp.2018}.

Here we revisit how this fundamental process is modelled in simulations of
particle dynamics in strong EM fields.
In contrast to previous work, we employ a photon emission rate that is differential in
scattering angle as well as energy, thereby resolving the beaming of the radiation
around the emitting particle's instantaneous velocity.
As a result, the accuracy of simulations based on Monte Carlo implementation
of localized emission events~\cite{ridgers.jcp.2014,gonoskov.pre.2015} is
substantially improved, when benchmarked against exact QED predictions of nonlinear Compton
scattering~\cite{blackburn.pop.2018}.
Simulations in the multiphoton, quantum radiation reaction regime demonstrate
that including the beaming is important for accurate modelling of the emission of moderate-energy photons.
The consequent transverse recoil may be neglected in many
high-intensity scenarios, but is distinguishable in the increase in the
divergence of an electron beam that collides with a linearly polarized laser
pulse, for experimental parameters accessible with present-day technology.
Furthermore, employing an angularly resolved spectrum permits
quantitative estimation of the accuracy of the simulations
in the low-energy part of the photon spectrum, where interference, i.e. nonlocal,
effects become important.

In natural units $\hbar = c = 1$ (as used throughout),
the photon emission rate per unit proper time, energy $\omega'$, and polar and
azimuthal scattering angles $\theta$ and $\varphi$, is~\cite{baier.book}
	\begin{multline}
	W^{(3)} = \frac{\partial^3 W}{\partial u \partial z \partial\varphi} =
		\frac{\alpha m}{3\sqrt{3}\pi^2 \chi}
		\frac{u}{(1+u)^3}
	\\
		\times \left\{
			z^{2/3} [1 + (1+u)^2] - (1 + u)
		\right\}
		K_{1/3}\!\left( \frac{2 u z}{3 \chi} \right),
	\label{eq:EmissionRate}
	\end{multline}
where $\alpha = e^2/(4\pi)$ is the fine structure constant, $e$ is the elementary
charge, $m$ is the electron mass, $u = \omega' / (\gamma m - \omega')$,
$z = [2\gamma^2(1 - \beta\cos\theta)]^{3/2}$,
and $K$ is a modified Bessel function of the second kind.
The spectrum is controlled by the electron's Lorentz factor $\gamma$ (velocity $\beta$)
and quantum nonlinearity parameter $\chi = e\abs{F_{\mu\nu} p^\nu}/m^3$.
Here $F$ is the EM field tensor and $p$ is the electron momentum.
The parameter $\chi$ may be interpreted as the ratio of the rest-frame electric field strength to
that of the critical field of QED $\Ecrit = m^2/e$~\cite{sauter.zp.1931,schwinger.pr.1951},
or as the magnitude of the proper acceleration in natural units.

The radiation is strongly beamed around the particle's instantaneous velocity
if the particle is ultrarelativistic~\cite{ritus.jslr.1985,seipt.prl.2017}.
The mean square angle of the power spectrum, $\avg{\theta^2} =
\int\! \theta^2 \omega' W^{(3)} \rmd{u}\rmd{z}\rmd{\varphi}
/ \int\! \omega' W^{(3)} \rmd{u}\rmd{z}\rmd{\varphi}$,
is $\avg{\theta^2} \simeq 5/(4\gamma^2) \ll 1$ in the classical limit
$\chi \ll 1$. It is larger in the quantum regime, growing as
$\avg{\theta^2} \simeq 1.76\gamma^{-2}\chi^{2/3}$ for $\chi \gg 1$, but still small.
This justifies the standard approximation used in simulation codes that photons are
emitted parallel to the particle momentum~\cite{ridgers.jcp.2014,gonoskov.pre.2015}.
Nevertheless, its inclusion is warranted because the degree of beaming depends
on the photon energy as well as the electron energy.
The mean square angle at fixed photon energy $\omega' = \gamma m u / (1+u)$,
$\avg{\theta^2(\omega')} = \int\! \theta^2 W^{(3)} \rmd{z}\rmd{\varphi}
/ \int\! W^{(3)} \rmd{z}\rmd{\varphi}$,
is, to leading order in $\chi/u$:
	\begin{equation}
	\gamma^2 \avg{\theta^2(\omega')} =
		\begin{cases}
		\frac{\Gamma(4/3)}{\Gamma(2/3)} \left( 3\chi / u \right)^{2/3}, & \chi / u\gg 1, \\
		\chi/u, & \chi / u\ll 1
		\end{cases}
	\label{eq:EnergyResolvedAngle}
	\end{equation}
The lower the photon energy, the larger its emission angle:
note that for $\omega' \ll \gamma m$, $u \simeq \omega'/(\gamma m)$.

We have implemented a Monte Carlo algorithm that samples the triple-differential
spectrum into a particle-tracking code, as an alternative to the standard method
in which only the energy is sampled from the angularly integrated spectrum.
These discrete emission events occur stochastically along the particles' classical
trajectories; between them, the dynamics are determined by the Lorentz force alone.
The electron recoil on emission is fixed by the conservation of momentum.
This `semiclassical' approach to QED is appropriate if the normalized amplitude of
the field $a_0$ satisfies $a_0^3/\chi \gg 1$, such
that the formation lengths of the photons are much smaller than the timescale
of the external field~\cite{ritus.jslr.1985,dinu.prl.2016}
and emission rates for a `locally constant' field can be employed.

We first confirm this by comparing the results of simulations which include
the radiation beaming, with exact QED in \cref{sec:Comparison}.
We propose a conceptually simple way to
estimate the magnitude of the error made by the `locally constant field'
emission rate used in simulations.
Then in \cref{sec:RadiationSpectrum,sec:ElectronDynamics}
we predict the beaming's effect on the radiation spectrum and electron dynamics
in experimentally relevant scenarios, where multiple photon emissions
and the spatiotemporal structure of the focussing laser field are taken into account.

\section{Improved agreement with exact QED}
\label{sec:Comparison}

Sampling the angularly resolved emission spectrum leads to substantially improved
agreement with exact QED results. The interaction we consider is single nonlinear
Compton scattering, i.e. the emission of one and only one photon by an electron in
an intense, pulsed plane EM wave.
The field tensor for the pulse is
$e F_{\mu\nu} = m a_0 \sum_i k_{[\mu} \varepsilon^i_{\nu]} \frac{\rmd \psi_i}{\rmd\phi}$,
where $(\psi_1,\psi_2) = (\cos\phi,\delta\sin\phi) \cos^2[\phi/(4\sigma)]$ for
$\abs{\phi} \leq 2 \pi \sigma$, $\phi$ is the phase, $\delta = 1$ and 0 for
circular and linear polarization, $k$ is the wavevector, and $\varepsilon_{1,2}$
are the polarization vectors along $x$ and $y$ respectively.

The one-photon emission probability is calculated in the framework of
strong-field QED, which accounts for the interaction with the background
electromagnetic field to all orders in $a_0$~\cite{furry.pr.1951,seipt.pra.2011,seipt.jpp.2016}.
The total probability,
which can exceed unity, is interpreted as the mean number of emissions
$N_\gamma$~\cite{ritus.jslr.1985,dipiazza.prl.2010}. As our Monte Carlo simulations allow
for the emission of an arbitrary number of photons, equivalent results are
obtained by post-selection~\cite{blackburn.pop.2018}:
photon spectra are generated statistically using only those simulated
collisions in which exactly one photon is emitted, and then rescaled to
have integral equal to the mean number of emissions, as determined from the
full set of collisions.

	\begin{figure}
	\subfloat[]{\label{fig:comp-a}\includegraphics[width=\linewidth]{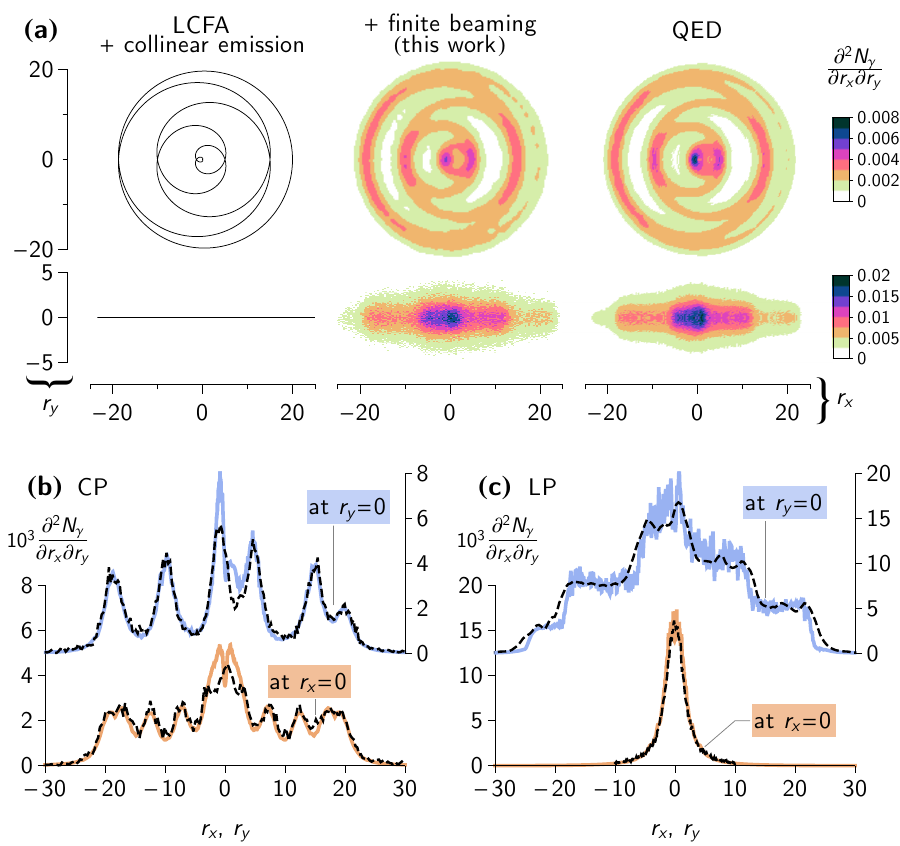}}
	\subfloat[]{\label{fig:comp-b}}
	\subfloat[]{\label{fig:comp-c}}
	\caption{
		Agreement with exact QED is improved when simulations using
		localized emission rates include the finite beaming of the radiation.
		(a) Differential probability that an electron emits a single photon
		with normalized perpendicular momenta $r_{x,y}$ in a circularly
		(CP) or linearly polarized (LP) EM wave.
		(b,c) Lineouts along $r_{x,y} = 0$:
		results from QED (solid colours) and simulations that include
		finite 	beaming (black, dashed).
		}
	\label{fig:Comparison}
	\end{figure}

A comparison between exact QED results and simulations that do and do not
include the finite beaming of the radiation is presented in \cref{fig:Comparison}.
Results are given in terms of the emitted photon's normalized perpendicular
momenta $m r_{x,y} = k'_{x,y} (k.p_0 / k.k')$, where $p_0$ is the initial
electron momentum.
The $r_x$-$r_y$ spectrum is effectively the angular profile
of the emitted radiation if $\gamma_0 \gg a_0 \gg 1$, as $r_{x,y} \simeq
\gamma_0 \theta_{x,y}$ for $\tan\theta_{x,y} = k'_{x,y}/(-k'_z)$.
We consider two examples: an electron with $\gamma_0 = 3000$
collides with a circularly polarized pulse with $a_0 = 20$; and
an electron with $\gamma_0 = 1\times10^4$ collides with a linearly polarized
pulse with $a_0 = 25$. $\sigma = 3$ and the central frequency
$\omega = k^0 = 1.55$~eV in both cases.

If the finite beaming is neglected, the calculated photon spectrum collapses onto a
curve that traces the electron trajectory: $m r_{x,y} = p_{x,y}(\phi)$.
This causes the angular spread of the radiation to be significantly
underestimated~\cite{blackburn.pop.2018}.
By contrast, when the finite beaming is included [central column of
\cref{fig:comp-a}], we obtain excellent agreement with the QED results [right column of
\cref{fig:comp-a}].
The structure of the angular profile is reproduced in both the circularly
and linearly polarized cases, as is shown by the lineouts along the axes
$r_x = 0$ and $r_y = 0$ in \cref{fig:comp-b,fig:comp-c}.

Models based on localized emission, as ours is, are accurate for photons
with energies $\omega'/(\gamma m) \gtrsim \chi/a_0^3$. Low-energy photons,
or those that are emitted in low-intensity regions of the pulse, have long
formation lengths and interference effects
tend to suppress their emission~\cite{harvey.pra.2015}.
Hence we observe discrepancies near $r_x = r_y = 0$ in the circularly
polarized case, because photons in this region originate from the pulse head
and tail where the local value of $a_0 \not\gg 1$.
Similarly, the spectrum in the linearly polarized case is broader in the $r_x$
direction near the turning points $\partial_\phi r_x = 0$, where the local
field vanishes.

	\begin{figure}
	\subfloat[]{\label{fig:flc-a}\includegraphics[width=\linewidth]{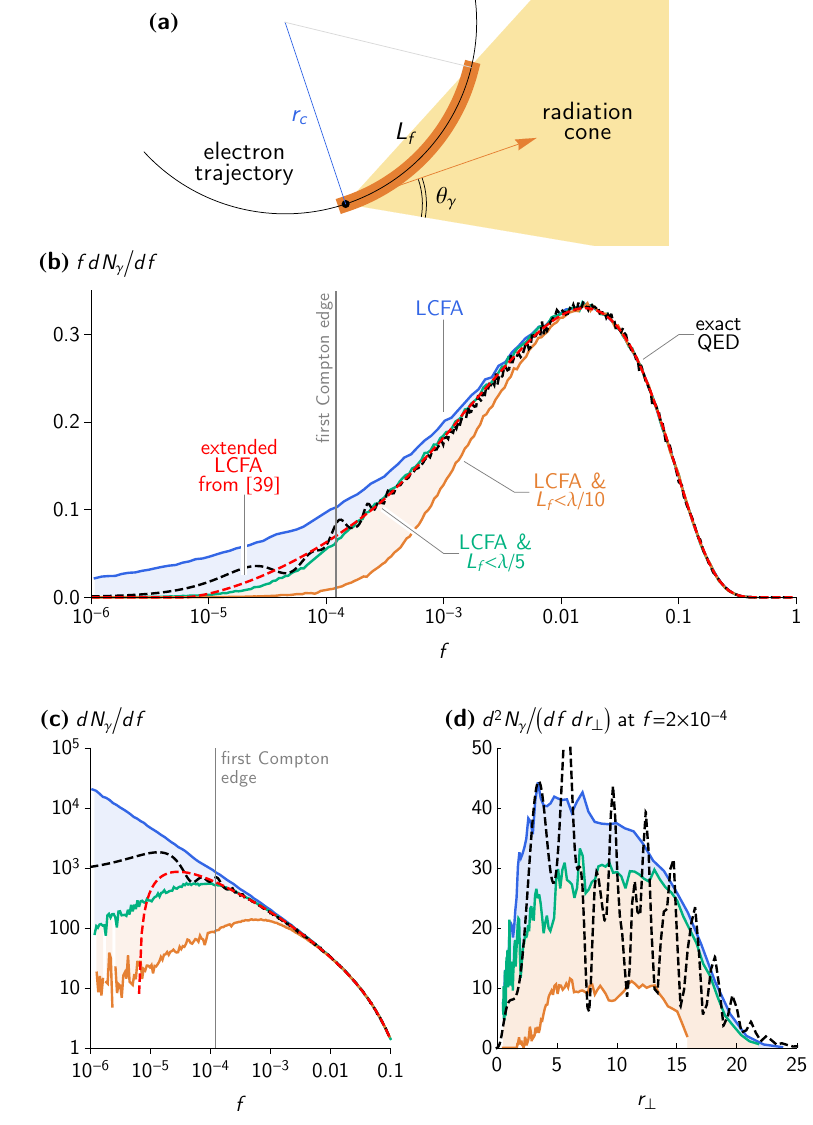}}
	\subfloat[]{\label{fig:flc-b}}
	\subfloat[]{\label{fig:flc-c}}
	\subfloat[]{\label{fig:flc-d}}
	\caption{
	    Estimating the error made in neglecting the finite formation
	    length of the emitted photon, at $a_0 = 10$:
	    (a) the relation between the formation length and the
	    angle at which the photon is emitted;
	    (b,c,d) comparison of results from exact QED (black, dashed)
	    and simulations where only photons with formation lengths
	    $L_f < \lambda/10$ (orange), $\lambda/5$ (green) or infinity (blue)
	    are emitted.
	    In (b,c) we also show the result using the ``extended'' LCFA
	    rates presented in \cite{ilderton.pra.2019} (red, dashed).
	    In (d) we have chosen a value of $r_\perp$ that lies in the region
	    where the validity of the LCFA is questionable.
		}
	\label{fig:FormationLength}
	\end{figure}

The finite formation length of the photon is a significant potential source
of error in simulations based on localized emission rates~\cite{blackburn.pop.2018}.
If this length is comparable to the spatial scale of variation of the
background field, nonlocal effects such as quantum interference become important.
We now discuss how sampling the angularly resolved photon spectrum allows the
magnitude of such effects to be estimated. Observe that, in the classical picture,
the formation length of a photon emitted at angle $\theta$ to the electron
instantaneous momentum is the distance travelled by the electron before it has
separated from the photon by at least $\theta$ (see \cref{fig:flc-a}).
This distance may be estimated locally as $L_f \simeq 2 r_c \theta$, where
the instantaneous radius of curvature of the electron trajectory
    \begin{align}
    r_c =
        \frac{\gamma^2-1}
        {\sqrt{m^2\chi^2 - m^2 (\vec{E} + \vec{v}\times\vec{B})^2/\Ecrit^2} }
    \end{align}
can be calculated using the Frenet-Serret formalism, assuming that only
electromagnetic forces are acting on the electron~\cite{seipt.ppcf.2019}.
For all practical purposes, the curvature radius can be approximated as
$r_c \simeq \gamma^2 / (m \chi)$, as is done in this paper.
$L_f$ can then be calculated for each simulated photon on emission, using the sampled value of
the angle $\theta$, and if it exceeds a specified maximum value, the photon
is not emitted.

Note that, because photons are only ever removed, this procedure does not
account for constructive interference effects that could enhance photon emission.
However, as it has been shown that the LCFA tends to lead to overestimation of
the low-energy part of the spectrum~\cite{harvey.pra.2015,dipiazza.pra.2018,blackburn.pop.2018},
comparing the results from simulations with and without this formation length
check provides a conceptually simple way to estimate the accuracy of the spectra predicted.

An example of this procedure is shown in \cref{fig:FormationLength},
which gives spectra that are differential in $f = k.k' / k.p_0$, the
lightfront-momentum transfer fraction, and $r_\perp = (r_x^2 + r_y^2)^{1/2}$,
for the photon emitted in the collision of electron with $\gamma_0 = 1000$
and a circularly polarized laser pulse with $a_0 = 10$ and $\sigma = 3$.
All three simulations include the finite beaming of the radiation,
but take different values of the maximum permitted formation length.
Observe that the spectra without a maximum (blue lines) and $L_f < \lambda/10$
(orange lines) bracket the exact QED result; the difference between the
two illustrates the expected accuracy of the LCFA.

\Cref{fig:flc-d} shows the double-differential spectrum at constant
$f = 2\times 10^{-4}$, which lies in the region $f \lesssim 2 \chi/a_0^3$,
where this accuracy is weakest.
The estimated error is large, warning that substantial interference
effects are expected, as visible in the exact QED result.
In fact, the best agreement is obtained with a formation
length cutoff of $\lambda/5$ (green lines in \cref{fig:FormationLength}),
which lies in between the two extreme cases.
It is similar to the result of a simulation using
the ``extended'' photon emission rates derived by \citet{ilderton.pra.2019}.
As this approach is based on gradient corrections to the LCFA, two
filters are necessary: one for the correction, which is
activated only for $a(\phi) > c = \pi/2$, and a global filter ensuring positivity of the rate.
Note that the extended rates are presented only in their angularly
integrated form and thus we cannot compare the angularly resolved
spectra shown in \cref{fig:flc-d}. We could obtain one by assuming
collinear emission, but it would have a hard cutoff at
$r_\perp = a_0 = 10$, which is not consistent with exact QED~\cite{blackburn.pop.2018}.

The procedure we have outlined uses only \emph{local} quantities ($\chi, \gamma$)
to estimate the formation length and it is therefore agnostic as to the specific structure
of the background field. However, if we explicitly choose this to be
an EM wave, where $\chi = 2 a_0 \gamma \omega / m$, and take
$\theta \simeq 1/\gamma$ as representative of the whole photon spectrum,
we recover the well-known result that $L_f \simeq 1/(a_0 \omega)$~\cite{ritus.jslr.1985}.
On the other hand, using \cref{eq:EnergyResolvedAngle} gives how the formation
length depends on the photon energy $\omega'$:
    \begin{equation}
    L_f \simeq
        \frac{\chi^{1/3}}{a_0 \omega}
        \left( \frac{\gamma m - \omega'}{\omega'} \right)^{1/3}.
    \label{eq:FormationLength}
    \end{equation}
This is consistent with results of \citet{dipiazza.pra.2018}.
No matter how large $a_0$ is, photons with sufficiently low energy
can have formation lengths comparable to the laser wavelength.
Our approach optionally excludes such photons on physical grounds,
putting error bars on theoretical predictions.
This is complementary to the use of corrected LCFA
rates~\cite{dipiazza.pra.2018,dipiazza.pra.2019,ilderton.pra.2019,king.arxiv.2019},
which aim to reduce the error rather than estimate its magnitude.

Thus far we have considered only the emission of a single
photon, as this can be calculated within QED and so
benchmarking of the angularly resolved LCFA rate \cref{eq:EmissionRate}
is possible.
We now turn to the effect of the radiation beaming on the
photon and electron spectra in more realistic scenarios,
where we allow for multiple photon emission and spatiotemporal
structure in both the laser pulse and electron beam.

\section{Broadening of the radiation angular profile}
\label{sec:RadiationSpectrum}

In a head-on collision with an EM wave that is linearly polarized along $x$,
neglect of the finite emission angle means that all photons have $r_y = 0$,
confining the radiation emitted by an initially divergence-free electron
beam to the laser polarization plane. In reality, photons are
emitted with $r_y \neq 0$. Thus as the initial divergence of the electron
beam is reduced to zero, the photon divergence in the perpendicular
direction (along $y$) saturates at a non-zero value.

This floor on the final divergence can be estimated analytically in the limit
$\chi \ll 1$, where the mean square polar angle
of the instantaneous power spectrum is $\avg{\theta^2} = 5/(4\gamma^2)$.
The total variance of the radiation angular profile in the $y$-direction,
$\delta_\gamma^2$, after the electron has passed through a pulsed
plane wave, is obtained by integrating $\frac{1}{2} \avg{\theta^2}$ over the pulse
temporal profile. Thus we have $\delta_\gamma^2 =
\int\! \frac{1}{2} \avg{\theta^2} \mathcal{P} \,\rmd\phi / \int\! \mathcal{P} \,\rmd\phi$,
where $\mathcal{P} = \alpha m^2 \chi^2 / (3\omega) \propto [\gamma(\phi) g(\phi)]^2$ gives
the instantaneous radiated power (per unit phase), $g(\phi)$ is the pulse temporal envelope,
and $\gamma(\phi)$ is the electron Lorentz factor as a function of phase $\phi$.
We obtain the latter by solution of the Landau-Lifshitz equation~\cite{landau.lifshitz},
which accounts for the deceleration due to classical radiation reaction.
Assuming that $g(\phi)$ is slowly varying, this gives
$\gamma(\phi) \simeq \gamma_0 / [1 + 2\alpha a_0^2 \gamma_0 \omega \mathcal{I}(\phi) / (3m)]$,
where $\mathcal{I}(\phi) = \int_{-\infty}^\phi g^2(\psi)\,\rmd\psi$.
Therefore
	\begin{align}
	\delta_\gamma^2 &=
		\delta_0^2
		+ \frac{5 (1 + \mathcal{R})}{8 \gamma_0^2},
	&
	\mathcal{R} &=
		\frac{2\alpha a_0^2 \gamma_0 \omega}{3 m}
		\int_{-\infty}^\infty\! g^2(\phi) \,\rmd\phi,
	\label{eq:PhotonDivergence}
	\end{align}
where $\delta_0$ is the initial divergence of the electron beam.
If the intensity profile $g^2(\phi)$ is a Gaussian with FWHM duration $\tau$,
$\int_{-\infty}^\infty g^2(\phi) \,\rmd\phi = \omega \tau \sqrt{\pi/(4\ln2)}$.

We compare this prediction to the results of 3D simulations of laser-electron
collisions. In contrast to our comparison with exact QED in \cref{sec:Comparison}, these
simulations account for \emph{multiphoton} radiation-reaction effects
as well as the spatiotemporal structure of the electron beam and focussed laser pulse.
The former is initialized with mean energy 500~MeV and root-mean-square (rms)
energy spread 50 MeV, divergence $\delta_0 = 0.5$~mrad and
size $\rho = 10~\micron$.
This corresponds to a normalized transverse emittance of
$\epsilon_\perp = [\avg{y^2}\avg{p_y^2/m^2}]^{1/2} = 5.0~\text{mm}\,\text{mrad}$.
Much smaller emittances have already been measured in laser-wakefield
accelerators~\cite{plateau.prl.2012,weingartner.prstab.2012}.
The laser pulse has wavelength $\lambda = 0.8~\micron$, duration 30~fs,
is linearly polarized along $x$, and is focussed
to a spot of size $\waist = 2.5~\micron$ and peak intensity $2\times10^{21}~\Wcmsqd$.
The fields in our simulations are calculated to fourth order in the
diffraction angle~\cite{salamin.apb.2007}.

	\begin{figure}
	\subfloat[]{\label{fig:ps-a}\includegraphics[width=\linewidth]{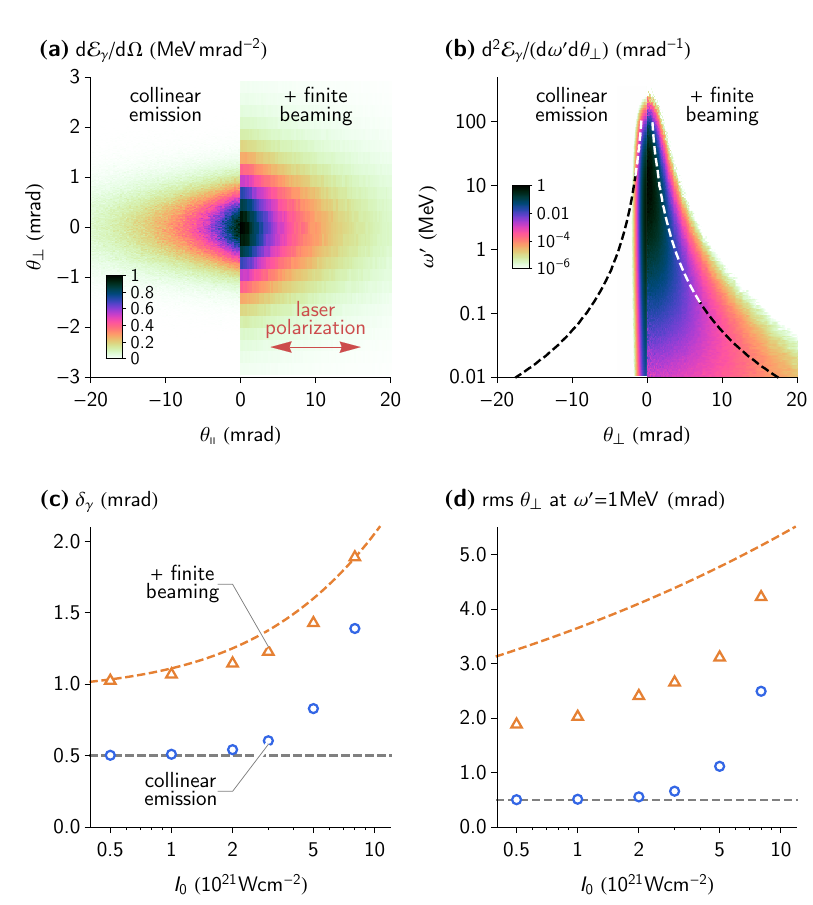}}
	\subfloat[]{\label{fig:ps-b}}
	\subfloat[]{\label{fig:ps-c}}
	\subfloat[]{\label{fig:ps-d}}
	\caption{
		Effect of the radiation beaming on the angularly resolved
		photon spectrum,
		in the collision of a 500-MeV electron beam and a
		linearly polarized laser pulse with peak intensity $I_0$.
		Density maps (colour scale, normalized to respective maxima)
		of (a) the energy radiated per unit solid angle
		and (b) the energy radiated per unit frequency
		and angle, both
		at $I_0 = 2\times 10^{21}~\Wcmsqd$.
		The divergence in the $y$-direction (c) of the total spectrum
		and (d) at fixed frequency $\omega' = 1$~MeV:
		simulation results (points),
		theoretical predictions of
		\cref{eq:PhotonDivergence,eq:EnergyResolvedAngle}
		(orange, dashed)
		and the initial beam divergence (grey, dashed).
		}
	\label{fig:PhotonSpectra}
	\end{figure}

The photon spectra for this scenario, resolved in $\theta_x$ (the angle in
the plane of polarization) and $\theta_y$ (the perpendicular angle),
are shown in \cref{fig:ps-a,fig:ps-b}; they are clearly broader in $\theta_y$
when the beaming of the radiation is included.
This demonstrates that the increase shown in the upper panel of \cref{fig:Comparison}
can survive more realistic interaction parameters.
Furthermore, \cref{fig:ps-b} shows that the angular spread
increases as the photon energy is lowered, whereas the entirety of the
radiation is confined to $\theta_y \lesssim 3\delta_0$ if emission is
assumed to be collinear.

\Cref{fig:ps-c,fig:ps-d} give the energy-weighted rms
$\theta_y$ of all photons, and only those photons with $\omega' = 1$~MeV,
as a function of peak intensity, with all other parameters fixed.
Both are in reasonable agreement with our theoretical predictions
\cref{eq:PhotonDivergence,eq:EnergyResolvedAngle}, setting
$\chi = 2\gamma_0 a_0 \omega /m$ and $\gamma = \gamma_0$ in the former.
Note that it is possible for $\delta_\gamma > \delta_0$ even if emission
is assumed to be collinear, because the decelerated electrons are
ponderomotively expelled from the focal spot in both the $x$- and $y$-directions.
In principle, the radiation beaming is also evident in the plane of polarization.
However, if $a_0 \gg 1$, the angular extent of the radiation in this direction is
dominated by the $a_0/\gamma$ contribution of the electron's oscillation.

The range of photon energies where inclusion of the beaming is essential
can be estimated as the range for which the typical emission angle
is between two and ten times the global average $\sim 1/\gamma$.
Using our earlier result, \cref{eq:EnergyResolvedAngle},
this is $\chi/870 \lesssim u \lesssim \chi/7$, where $\omega'/(\gamma m) = u/(1+u)$.
For the parameters used in \cref{fig:PhotonSpectra}, this corresponds
to photons with energies from 0.1 to a few MeV.
Even though they individually contribute little to the total energy loss,
such photons are emitted in far greater numbers than their higher energy
counterparts.
As discussed in \cref{sec:Comparison}, simulations based on the LCFA
tend to overestimate the yield of low-energy photons; thus we validate
the results shown in \cref{fig:PhotonSpectra} against simulations in
which photons with formation length $L_f > \lambda/10$ are discarded.
This reduces the number of 1-MeV photons by 40\%, but the additional broadening
at this energy due to the finite beaming (\cref{fig:ps-b}) and the total energy
radiated per unit solid angle (\cref{fig:ps-a}) are unchanged.
Similarly, the angular widths given in \cref{fig:ps-b,fig:ps-d} are unchanged to within 5\%.

	\begin{figure}
	\includegraphics[width=\linewidth]{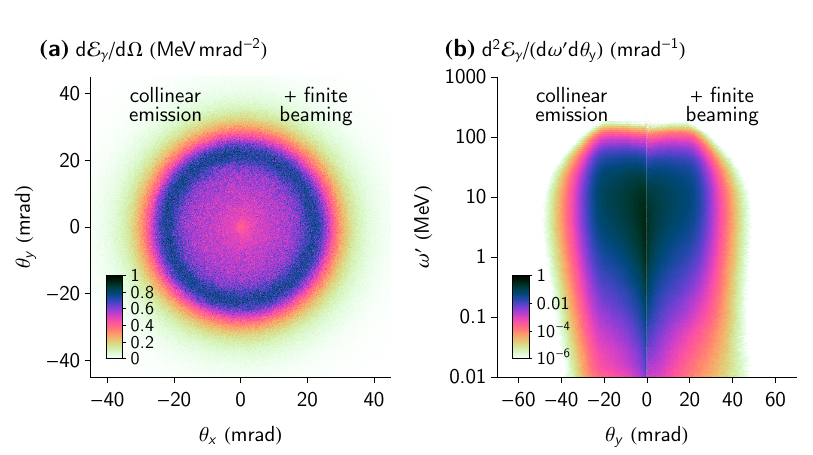}
	\caption{
		The effect of the radiation beaming on the angularly resolved
		photon spectrum is much weaker when the laser is circularly
		polarized:
		density maps (colour scale, normalized to respective maxima)
		of (a) the energy radiated per unit solid angle
		and (b) the energy radiated per unit frequency and angle 
		in the collision of a 500-MeV electron beam and a
		circularly polarized laser pulse with peak intensity $2\times10^{21}~\Wcmsqd$.
		}
	\label{fig:Circular}
	\end{figure}

It is important to note that, if the laser is circularly rather than
linearly polarized, there is no distinction between the two directions
perpendicular to the wavevector.
The electrons oscillate in $x$ and $y$ and therefore the radiation
has a finite angular spread in both directions, even if the initial
electron divergence is reduced to zero and the finite beaming is neglected.
This is shown in \cref{fig:Circular}, where we compare the energy emitted
per unit solid angle by a 500-MeV electron beam colliding with a circularly
polarized, plane-wave, laser pulse, with peak intensity $2\times10^{21}~\Wcmsqd$,
wavelength $0.8~\micron$, and duration 30~fs. (The electron beam has
rms energy spread 50~MeV and divergence $\delta_0 = 0.5$~mrad.)
Comparing \cref{fig:Circular} to \cref{fig:PhotonSpectra}, it is clear
that a distinction between the two perpendicular directions is necessary
to observe finite beaming effects. We focus, therefore, on the case
of linear polarization.

\section{Quantum limit on the electron beam divergence}
\label{sec:ElectronDynamics}

We now turn to the consequences of non-collinear emission for the electron.
The conservation of momentum requires that if the photon is emitted at some
finite angle, a recoil $\Delta p$ is exerted on the emitting particle in the direction
perpendicular to its velocity.
$\Delta p = \omega' \sin\theta \simeq m u \sqrt{z^{2/3}-1}/(1+u)$
to leading order in $1/\gamma$; its mean value is $\avg{\Delta p}/m
\simeq 3\sqrt{3}\pi\chi/40$ for $\chi \ll 1$ and $0.264\chi^{1/3}$ for $\chi \gg 1$.
For the perpendicular component of the recoil to have a significant impact
on the dynamics, it should be comparable in size to the electron's
transverse momentum $p_x = m a_0$.
However, $\avg{\Delta p}/p_x \simeq 0.4\chi/a_0$ or $0.3\chi^{1/3}/a_0 \ll 1$
in almost all high-intensity scenarios of interest.
As such, it is safe to neglect the transverse recoil in models of quantum
radiation reaction, even though the emission probability vanishes for $\theta \to 0$
and therefore the recoil is never antiparallel to the instantaneous velocity.

Nevertheless, the effect of this transverse recoil can be visible in the collision
of an ultra-low emittance electron beam with a high-intensity, \emph{linearly}
polarized laser pulse. This is because, in a plane wave, the momentum in the
direction perpendicular to the polarization $p_y$ is preserved by the
Lorentz force; under classical radiation reaction, it can only
ever decrease. Concretely, the equations of motion for this scenario are
$\frac{\rmd}{\rmd\phi} (k.p) = -2\alpha m^2 \chi^2 / 3$ and
$\frac{\rmd}{\rmd\phi} [p_y / (k.p)] = 0$~\cite{landau.lifshitz}.
We have $p_y = p_{y,0} (k.p / k.p_0) \leq p_{y,0}$ by
$\frac{\rmd}{\rmd\phi} (k.p) \leq 0$, where the equality applies in the absence
of radiation reaction. If $p_y = 0$ initially, it remains so.
This is no longer the case when the transverse recoil is included.

Provided that radiation losses are not too large, the electron emerges from
the laser field with $k.p \simeq 2\omega\abs{p_z}$. Therefore the distribution
of $\tan \theta_y = p_y/\abs{p_z} \simeq 2\omega (p_y / k.p)$
is unchanged under classical radiation reaction. It is unchanged under quantum
radiation reaction only if collinear emission is assumed.
Including the finite emission angle and associated transverse recoil,
by contrast, leads to an increase in the out-of-plane divergence.
As the initial divergence of the electron beam is reduced to zero, the final
divergence $\delta_e = \avg{\theta_y^2}^{1/2}$ saturates at a non-zero value.

This lower bound on the divergence is a pure quantum effect,
arising from the finite number of emissions.
This phenomenon will occur not only in an ultraintense laser, as considered here,
but also in a static magnetic field.
In principle, the transverse recoil sets a lower bound on the emittance
of an electron beam in a storage ring, in the direction parallel to the
magnetic field~\cite{raubenheimer.slac.1991};
however, this limit is typically four orders of magnitude
smaller than the emittance in the plane of the orbit, and in practice
is dominated by magnet alignment errors and other deviations.

To estimate the final divergence, we assume that the electron performs a random walk in $\theta_y$, so
$\delta_e^2 = \int\!\frac{1}{2}\avg{\theta_e^2} W_\phi \,\rmd\phi$, where
the electron polar scattering angle $\theta_e \simeq u \sqrt{z^{2/3}-1}/\gamma$
and $W_\phi$ is the instantaneous rate of photon emission per unit phase.
In the limit $\chi_e \ll 1$, $\avg{\theta_e^2} \simeq 13\chi_e^2/(30\gamma^2)$
and $W_\phi \simeq 5\alpha m \chi_e / (4 \sqrt{3} \gamma \omega)$.
Assuming that the temporal profile $g(\phi)$ is slowly varying, we find
	\begin{equation}
	\delta_e^2 = \delta_0^2 +
		\frac{26 \sqrt{3} \alpha}{27\pi}
		\frac{a_0^3 \omega^2}{m^2}
		\int_{-\infty}^\infty\! g^3(\phi) \,\rmd\phi,
	\label{eq:TrrDivergence}
	\end{equation}
where $\delta_0$ is the initial divergence of the electron beam.
If the intensity profile is a Gaussian with peak $I_0$ and FWHM duration $\tau$,
$\int_{-\infty}^\infty\! g^3(\phi) \,\rmd\phi = \omega \tau \sqrt{\pi/(6\ln2)}$ and
$\delta_e[\text{mrad}] \simeq 0.086 I_0^{3/4}[10^{21}\,\Wcmsqd] \tau^{1/2}[10\,\fs]$.

	\begin{figure}
	\subfloat[]{\label{fig:es-a}\includegraphics[width=\linewidth]{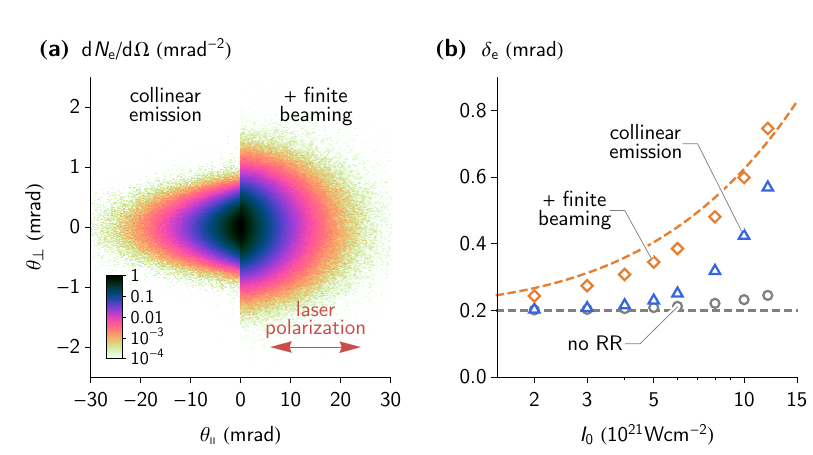}}
	\subfloat[]{\label{fig:es-b}}
	\caption{
		Effect of the transverse recoil on the electron angular distribution,
		in the collision of a 500-MeV electron beam and a
		linearly polarized laser pulse with peak intensity $I_0$:
		(a) $\theta_x$-$\theta_y$
		distribution at $I_0 = 5\times 10^{21}~\Wcmsqd$;
		(b) rms $\theta_y$ from simulations (points),
		its initial value (grey, dashed), and that predicted by
		\cref{eq:TrrDivergence} (orange, dashed).
		}
	\label{fig:ElectronSignatures}
	\end{figure}

We now compare this prediction to the results of 3D simulations of
laser-electron collisions. It is essential to account for multidimensional effects,
because there is a ponderomotive contribution to the electron
deflection~\cite{quesnel.pre.1998}, which is enhanced by energy losses to
radiation emission. To mitigate this competing source of divergence increase,
we consider collisions with frequency-doubled laser pulses that are focussed
to relatively large spot sizes.
This exploits the fact that the ponderomotive force, and so the divergence
it induces, are proportional to the gradient of the squared vector potential,
$\nabla a^2(x,y) \propto I_0 \lambda^2 \rho / w_0^2$,
whereas the increase in divergence due to finite beaming $\delta_e \propto I_0^{3/4}$
depends only on intensity.
The electron beam is initialized with mean energy 500~MeV, energy spread
100 MeV, divergence $\delta_0 = 0.2$~mrad and size $\rho = 1.0~\micron$ (all rms),
which corresponds to $\epsilon_\perp = 0.2~\text{mm}\,\text{mrad}$~\cite{plateau.prl.2012,weingartner.prstab.2012}.
The frequency-doubled laser pulse has wavelength $0.4~\micron$, duration 15~fs,
and is focussed to a spot of size $\waist = 5~\micron$ and peak intensity
$5\times10^{21}~\Wcmsqd$.

The electron angular distributions for this particular configuration are shown
in \cref{fig:es-a}; the variation of the rms angle with peak
intensity (all other parameters unchanged) is shown in \cref{fig:es-b}.
We see that the transverse recoil leads to a greater increase in the
perpendicular divergence than quantum radiation reaction alone (i.e., if
emission and recoil are assumed to be collinear with the electron initial
momentum). The rms perpendicular angle obtained in the simulations
agrees well with \cref{eq:TrrDivergence}.
These results are unchanged if the simulations are rerun with a 
maximum permitted photon formation length of $L_f = \lambda/10$,
using the procedure given in \cref{sec:Comparison}.
This confirms that the beaming of the radiation is important for
photons that are sufficiently energetic to affect the electron,
unlike interference effects~\cite{blackburn.pop.2018}.
The challenge in realizing such measurements is the
high degree of control required over both electron beam and laser
pulse. Furthermore, we cannot simply increase the peak
intensity to yield a larger value of $\delta_e$, as \cref{eq:TrrDivergence}
suggests, because this enhances radiative energy losses and so
the ponderomotive deflection that masks the relevant signal.
We will explore such effects in detail elsewhere.

\section{Conclusions}

The radiation emitted by ultrarelativistic charged particles
is strongly beamed in the direction parallel to the particle velocity.
Despite the smallness of the opening angle, we have shown
that implementation of a photon emission rate that is resolved in
scattering angle as well as energy is necessary for accurate
simulations of radiation generation in the quantum regime.
The finite beaming is particularly important for moderate-energy photons,
which are emitted into a broader range of angles.

The finite emission angle means that there is a component of the recoil
that is perpendicular to the unperturbed momentum. While negligible
in many high-intensity scenarios of interest,
we have shown that this transverse recoil
leads to a lower bound on the divergence of the electron beam
in the direction perpendicular to the plane defined by the unperturbed
momentum and the force of the external electromagnetic field.
The increase in the out-of-plane momentum is a purely quantum effect, even
though radiation beaming is a feature of the classical theory as well.
This is because the number of emissions $N_\gamma \to \infty$ in the limit
$\hbar \to 0$, which averages the recoil over the arbitrary
azimuthal angle. In the quantum regime, the number of emissions is finite
and therefore the change in transverse momentum is not completely compensated.
The consequent increase in the electron beam divergence is a signature
of radiation reaction dynamics that go beyond the stochastic effects
previously considered~\cite{green.prl.2014,vranic.njp.2016,li.pra.2018}.

Beyond the interaction with a single laser pulse examined here,
it is possible that the transverse recoil affects cascade development in an
EM standing wave~\cite{bell.prl.2008,fedotov.prl.2010,bulanov.prl.2010b},
because it would displace electrons from electric-field
antinodes~\cite{king.prd.2016}, where the most energetic photons are emitted.
It might also seed plasma instabilities in dipole-wave-driven
cascades~\cite{efimenko.sr.2018}.

\begin{acknowledgments}
The authors acknowledge support from
the Knut and Alice Wallenberg Foundation (T.G.B., M.M.),
the US DOE under Contract No. DE-AC02-05CH11231 (S.S.B.),
the US ARO grant no.~W911NF-16-1-0044 (D.S.),
and the Swedish Research Council, grants 2013-4248 and 2016-03329 (M.M.).
Simulations were performed on resources provided by the Swedish National
Infrastructure for Computing (SNIC) at the High Performance Computing
Centre North (HPC2N).
\end{acknowledgments}

\bibliography{references}

\end{document}